\documentclass[10pt,twocolumn,twoside]{IEEEtran}
\usepackage{amsmath,amsfonts,amssymb,amsbsy,bm,paralist,theorem,ifthen,color}
\usepackage{graphicx,amssymb,amsmath}
\usepackage{subfigure,relsize,geometry}
\usepackage{theorem}
\usepackage{url}
\usepackage{algorithmic,algorithm}
\usepackage[square,comma, sort&compress, numbers]{natbib}
\usepackage{cases}
\usepackage{thmtools}

\DeclareMathOperator*{\st}{s.t.}

\newtheorem{lem}{Lemma}
\newtheorem{thm}{Theorem}

\newtheorem{cor}{Corollary}

\graphicspath{{fig/}} 

\begin{document}
\title{Energy-Efficient Non-Orthogonal Transmission under Reliability and Finite Blocklength Constraints}
\date\today\vspace{2mm}
\author{Yanqing Xu\IEEEauthorrefmark{1}, Chao Shen\IEEEauthorrefmark{1}, Tsung-Hui Chang\IEEEauthorrefmark{2}\IEEEauthorrefmark{3}, Shih-Chun Lin\IEEEauthorrefmark{4}, Yajun Zhao\IEEEauthorrefmark{5}, and Gang Zhu\IEEEauthorrefmark{1}\\
\IEEEauthorblockA{
\IEEEauthorrefmark{1}State Key Laboratory of Rail Traffic Control and Safety, Beijing Jiaotong University, Beijing, China\\
\IEEEauthorrefmark{2}School of Science and Engineering, The Chinese University of Hong Kong, Shenzhen, China\\
\IEEEauthorrefmark{3}Shenzhen Research Institute of Big Data, Shenzhen, China\\
\IEEEauthorrefmark{4}Department of ECE, National Taiwan University of Science and Technology, Taipei, Taiwan\\
\IEEEauthorrefmark{5}Algorithm Department, Wireless Product R\&D Institute, ZTE Corporation, Shenzhen, China\\
Email: \{xuyanqing, chaoshen, gzhu\}@bjtu.edu.cn, tsunghui.chang@ieee.org, sclin@mail.ntust.edu.tw, zhao.yajun1@zte.com.cn}\\
\thanks{ The work of Y. Xu and C. Shen were supported by the Fundamental Research Funds for the Central Universities (No. 2017YJS017), NSFC (No. 61501024, No. U1334202), the Fundamental Research Funds for the Central Universities (No. 2017JBM315), and the State Key Laboratory of Rail Traffic Control and Safety (No. RCS2016ZZ004). Tsung-Hui Chang was supported by NSFC, China, Grant No. 61571385 and CUHK(SZ) President Fund PF.01.000183.  Shih-Chun Lin was supported in part by the Ministry of Science and Technology, Taiwan, under Grant 104-2628-E-011-008-MY3}
}

\thispagestyle{empty}

\maketitle

\pagestyle{empty}  
\thispagestyle{empty} 

\begin{abstract}
This paper investigates an energy-efficient non-orthogonal transmission design problem for two downlink receivers that have strict reliability and finite blocklength (latency) constraints.
The Shannon capacity formula widely used in traditional designs needs the assumption of infinite blocklength and thus is no longer appropriate.
We adopt the newly finite blocklength coding capacity formula for explicitly specifying the trade-off between reliability and code blocklength.
However, conventional successive interference cancellation (SIC) may become infeasible due to heterogeneous blocklengths. We thus consider several scenarios with different channel conditions and with/without SIC. By carefully examining the problem structure, we present in closed-form the optimal power and code blocklength for energy-efficient transmissions. Simulation results provide interesting insights into conditions for which non-orthogonal transmission is more energy efficient than the orthogonal transmission such as TDMA.
\end{abstract}

\begin{IEEEkeywords}
  Ultra-reliable and low-latency communications (URLLC), finite blocklength codes, energy efficiency, non-orthogonal transmission.
\end{IEEEkeywords}

\vspace{-0mm}
\section{Introduction}
The vision of the 5G communication system promises to support ultra reliability and low latency for communications incurred by massive autonomous machines\cite{Popviski-2016-Procceding}. The requirement of such an ultra-reliable and low-latency communication (URLLC) system is drastically different from that of the 4G LTE, and is specified with no less than $1\!-\!10^{-6}$ reliability (i.e., $10^{-6}$ packet error probability), no longer than $1$ms latency and small packet size (e.g., $20$ bytes) \cite{urllc-ICCW}. Therefore, it requires new system architecture and signalling schemes to achieve URLLC, especially for multi-user communications.

To improve the reliability and the throughput, the non-orthogonal multiple access (NOMA) scheme, which allows multiple users to transmit simultaneously, has been considered in 5G system. Compared to the orthogonal multiple access (OMA), NOMA can exploit the channel diversity more efficiently via smart interference management techniques such as successive interference cancellation (SIC) \cite{Ding-Magazine}.  However, the existing NOMA designs cannot guarantee low latency especially when the coding blocklengths are short. This is because the current NOMA designs \cite{Ding-Magazine}\cite{XU-TSP-2017} are based on the classical Shannon capacity formula, which however is accurate only when the blocklength is infinitely long \cite{Book_Cover}. Recently, the capacity of finite blocklength coding (FBC) in AWGN channel has been characterized in \cite{2010-Polyasiki-TIT}. This new capacity formula explicitly characterizes the relationship between transmission rate, code blocklength and decoding reliability, thus particularly suitable for evaluating the performance of URLLC systems. The capacity of FBC has been successfully applied to the study of wireless communications with strict latency constraints, as in \cite{Xu-2016,Hu-2016-TWC,Sun-2017,Popovski-tcomm,Makki-tcomm,Ozcan-jsac}. For example, reference \cite{Xu-2016} considered the energy-efficient packet scheduling problem and showed that the classical Shannon capacity may significantly underestimate the energy under FBC. 
Reference \cite{Popovski-tcomm} considered a multi-user downlink scenario and proposed to group user messages for benefiting performance gain from long code transmissions.

In this paper, we consider optimal resource allocation for a two-user NOMA downlink with FBC. In particular, we formulate a new problem that aims to minimize the energy of the transmitter subject to heterogeneous reliability and latency constraints at receivers. Due to heterogeneous latency (blocklength) constraints, in contrast to conventional NOMA with homogeneous constraints, SIC may not always be feasible. Thus for different channel conditions, we will consider different interference management techniques according to whether SIC is feasible or not. Moreover, solving our optimization problem is challenging because the FBC capacity formula does not admit a closed-form expression for the energy function. With the aids of the implicit function theory \cite{Krantz_Parks02}, we identify the monotonicity of the energy function with respect to the blocklength. The optimal solutions of the considered problem under different channel conditions are then obtained.  Simulation results reveal that the NOMA transmission schemes are more energy efficient  than the OMA ones when receivers have similar latency constraints. However, the OMA scheme may become a better option if the receivers have markedly different latency requirements.

\section{System Model and Problem Formulations}
We consider a single-antenna NOMA downlink where a transmitter wants to send two private messages to two receivers respectively. According to the NOMA principle,
the transmitter encodes the $N_k$ message bits for receiver $k$ into a codeword with block length $m_k$ (symbols), $k=1,2$; and transmit the superposition of these two codewords to the receivers. The transmitted signal is then
$
\sqrt{p_1} x_{1} + \sqrt{p_2} x_{2}.
$
Here  $x_{1}$ and $x_{2}$ are the unit-power coded symbols of receiver $1$ and receiver $2$ respectively, and $p_1$ and $p_2$ are the transmission powers allocated to receiver $1$ and receiver $2$ respectively. The received signal for receiver $k$ is given by
\begin{align} \label{eq_downlink}
y_k = h_k (\sqrt{p_1} x_{1} + \sqrt{p_2} x_{2}) + n_k, \quad k = 1,2,
\end{align}
where $h_k \in \mathbb{C}$ and $n_k \sim \mathcal{CN}(0,\sigma_k^2)$ are the channel coefficient and noise at receiver $k$, respectively. Without loss of generality, we assume $\sigma_k^2=1, \forall k$. Different from the traditional downlink schemes \cite{Book_Cover}, \textit{strict latency constraint}s are imposed such that the codeword block length $m_k$ must be smaller than $D_k$ symbols (channel uses), $k=1,2$. Note that $D_k=\infty, \forall k$ in traditional downlink \cite{Book_Cover}. To cope with the new latency constraints, we adopt the FBC capacity formula in \cite{2010-Polyasiki-TIT} since the classical Shannon capacity formula is no longer appropriate.

\begin{figure}[!t]
	\centering
	\includegraphics[width=0.8\linewidth]{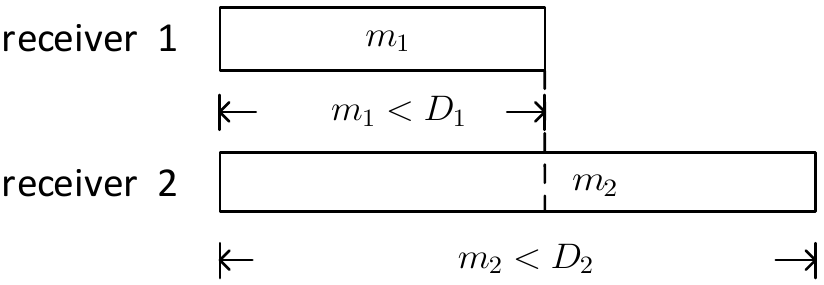}\\
	\caption{ Latency-constrained NOMA downlink, where the deadline $D_2$ of receiver $2$ is longer than that of receiver $1$. } \label{noma_large_D2}
	\vspace{-2mm}
\end{figure}

Besides the encoder, the conventional SIC based decoders in \cite{Book_Cover} also need to be re-designed due to the latency constraints. Note that our latency constraints can be heterogeneous such that $D_1 \neq D_2$. Thus unlike \cite{Book_Cover}, when $|h_1|>|h_2|$ and $D_1 < D_2$, receiver 1 \textit{may not} be able to decode receiver 2' message and cancel the corresponding interference. Also, the signal $y_2$ received at receiver 2 may not be a degraded (always worse) version of $y_1$. Thus one needs to design decoding strategies according to not only the channel gains $h_k$s but also the heterogeneous latency constraints $D_k$s. Without loss of generality, we assume $D_1 < D_2$ as in Figure \ref{noma_large_D2} and consider two cases in this paper, that is, $|h_1| \leq |h_2|$ and $|h_1| > |h_2|$. The corresponding energy optimization problems are detailed in the next subsection

\vspace{-2mm}
\subsection{Energy Minimization Problem when $|h_1| \leq |h_2|$} \label{subsec_largerh2}
Let us start from the case of $|h_1| \leq |h_2|$. Since $D_1 < D_2$, receiver 2 can apply SIC whereas receiver 1 cannot. Specifically, for receiver 1, interference symbol $x_2$ is treated as noise in \eqref{eq_downlink}. Thus the achievable rate under FBC is given by \cite{2010-Polyasiki-TIT}\cite{Xu-2016}
\begin{align}
\frac{N_1}{m_1} \!=\! \log_2(1 \!+\! \gamma_1) \!-\! \sqrt{\frac{1}{m_1}\left(1 \!-\! \frac{1}{(\gamma_1 \!+\! 1)^2}\right)} \frac{Q^{-1}(\epsilon_1)}{\ln2}, \label{4}
\end{align}
where $\gamma_1 = \frac{p_1 |h_1|^2}{p_2 |h_1|^2 + 1}$ is the received signal-to-interference-plus-noise ratio (SINR) for receiver $1$, $\epsilon_1$ is the predefined block error probability for receiver $1$, and $Q^{-1}(\cdot)$ is the inverse of the Gaussian Q-function \footnote{ Compared with the AWGN capacity upper-bound in \cite[equation (612)]{2010-Polyasiki-TIT}, achievable rate in \eqref{4} has loss within $\frac{\log(m_1)+\mathcal{O}(1)}{m_1}$}. By the principle of SIC, receiver $2$ would decode receiver 1's codeword with SINR $\frac{p_1 |h_2|^2}{p_2 |h_2|^2 + 1}$ in the first stage. Since $|h_1| \leq |h_2|$, the SINR value $\frac{p_1 |h_2|^2}{p_2 |h_2|^2 + 1}$ is higher than $\gamma_1$ and therefore the SIC can achieve the corresponding block error probability $\epsilon_{1}$. By successfully subtracting $x_1$ from $y_2$ in \eqref{eq_downlink} with probability $1-\epsilon_1$, receiver $2$ then decodes its private message and with probability $1-\epsilon_{2}$.
\begin{align}
\frac{N_2}{m_2} = \log_2(1 \!+\! \gamma_2) \!-\! \sqrt{\frac{1}{m_2}\left(1 \!-\! \frac{1}{(\gamma_2 \!+\! 1)^2}\right)} \frac{Q^{-1}(\epsilon_2)}{\ln2}, \label{5}
\end{align}
where $\gamma_2 = p_2 |h_2|^2$ and $\epsilon_2$ is the error probability conditioned on correct SIC. By considering the probability that SIC may be incorrect,  the overall decoding error probability of receiver 2 is given by $\bar{\epsilon}_2=\epsilon_1+(1-\epsilon_1)\epsilon_2$.

Based on the above models, the latency-constrained energy-efficient design problem when $|h_1|<|h_2|$ is formulated as
\begin{subequations}\label{P_NOMA1}
\begin{align}
\min_{\{m_k,p_k,\gamma_k\}_{k=1,2}} \quad & m_1 p_1 + m_2 p_2 \label{P_NOMA1 EF}  \\
\st \quad\quad\quad & F_k(m_k,\gamma_k) = 0, \quad \forall k = 1,2, \label{p1.1}\\
& \hat{m} \le m_k, \quad \forall k = 1,2,\label{p1.2}\\
& p_1 + p_2 \le P_{\rm max}, \quad \label{p1.4}\\
& 0 \le p_k, \quad \forall k = 1,2,\label{p1.5}\\
& m_k \le D_k, \quad \forall k = 1,2,\label{p1.3}\\
& \gamma_1 = \frac{p_1 |h_1|^2}{p_2 |h_1|^2 + 1}, \label{p1.6}\\
& \gamma_2 = p_2 |h_2|^2, \label{p1.7}
\end{align}
\end{subequations}
where \eqref{p1.3} are the latency constraints, and \eqref{p1.1} are the FBC constraints with
\begin{align}
F_k(m_k,\gamma_k) & \triangleq \sqrt{\frac{1}{m_k}\left(1 - \frac{1}{(\gamma_k + 1)^2}\right)} \frac{Q^{-1}(\epsilon_k)}{\ln2} \notag\\
&~~~~~~~~~~~~~~~~~~- \log_2(1 + \gamma_k) + \frac{N_k}{m_k} \label{transformation_1}.
\end{align}
Note that \eqref{p1.1} with $k=1$ corresponds to \eqref{4}, and  \eqref{p1.1} with $k=2$ corresponds to \eqref{5}. Constraints \eqref{p1.2} represents the minimum blocklength constraint for \eqref{p1.1} holding true \cite{2010-Polyasiki-TIT}\cite{Xu-2016} (typically $\hat{m}=100$), while \eqref{p1.4} and \eqref{p1.5} are the transmission power constraints.

Here we remark that solving problem \eqref{P_NOMA1} is challenging. In particular, the variables are coupled in the constraints in a non-linear and complex fashion. However, in upcoming Section \ref{subSec_solution1}, we will show how \eqref{P_NOMA1}  can be globally solved.

\vspace{-5mm}
\subsection{Energy Minimization Problem when $|h_1| > |h_2|$}
As aforementioned, unlike the case of $|h_1|\le |h_2|$, SIC may not be always valid when $|h_1| > |h_2|$ and $D_1 < D_2$. Thus we consider two scheduling policies as follows.\\
\noindent \textbf{B.1 Full latency for receiver 2:} In this case, we allow $m_1\leq m_2$ since $D_1 < D_2$. Therefore, receiver 1 may not be able to perform SIC, but instead treats $x_2$ as noise. Then the energy minimization problem is formulated as
\begin{subequations}\label{noma_case_2}
\begin{align}
\min_{\{m_k,p_k,\gamma_k\}} \quad & m_1 p_1 + m_2 p_2\\
\st \quad\quad & \eqref{p1.1}-\eqref{p1.5}, \notag \\
& m_k \le D_k, \quad \forall k = 1,2,\label{p3.5}\\
& \gamma_1 = \frac{p_1 |h_1|^2}{p_2 |h_1|^2 + 1}, \label{p3.6}\\
& \gamma_2 = \frac{p_2 |h_2|^2}{p_1 |h_2|^2 + 1}.\label{p3.7}
\end{align}
\end{subequations}

\noindent \textbf{B.2 Short latency for receiver 2:} In this case, we force
\[
m_2 \le m_1.
\]
Note that $m_1\leq D_1<D_2$, thus the original latency constraint $m_2 \leq D_2$ is automatically satisfied. Under the setting of $m_2 \le m_1$, SIC can be performed at receiver $1$ to completely remove the interference from receiver $2$. Then the energy minimization problem is formulated as
\begin{subequations}\label{noma_case_21}
\begin{align}
\min_{\{m_k,p_k,\gamma_k\}} \quad & m_1 p_1 + m_2 p_2\\
\st \quad \quad&\eqref{p1.1}-\eqref{p1.5}, \notag \\
& m_1 \le D_1, \label{p4.2}\\
& m_2 \le m_1,\\
& \gamma_1 = p_1 |h_1|^2, \label{p4.6}\\
& \gamma_2 = \frac{p_2 |h_2|^2}{p_1 |h_2|^2 + 1}.\label{p4.7}
\end{align}
\end{subequations}

The solutions of aforementioned two problems are given in Section \ref{subSec_solution2}. Here we point out that problem \eqref{noma_case_21} can yield a smaller energy than \eqref{noma_case_2} when the two deadlines $D_2$ and $D_1$ are close, thanks to the performance gain brought by SIC. However, when $D_2$ is significantly larger than $D_1$, formulation \eqref{noma_case_2} can become more energy efficient by benefiting from long code transmission.

\section{ Solutions of Energy-minimization Problems} \label{Sec_solution}
In this section, we present the solutions to the energy-minimization problems in \eqref{P_NOMA1}, \eqref{noma_case_2}, and \eqref{noma_case_21}.

\subsection{ Optimal Solutions for Problem \eqref{P_NOMA1}} \label{subSec_solution1}

By \eqref{p1.1} and by applying the implicit function theorem \cite{Krantz_Parks02},  there exist continuously differentiable implicit functions $\Gamma_k(\cdot)$ such that
\begin{equation}\label{eq_Gamma_P_MOMA}
\Gamma_k (m_k) = \gamma_k, \forall~k=1,2.
\end{equation}
From \eqref{p1.6} and \eqref{p1.7}, we can rewrite the target energy of \eqref{P_NOMA1 EF} as function of block length $m_k$s as
\begin{equation}\label{eq_P_MOMA_target}
\frac{m_1 \Gamma_1(m_1) \left(\Gamma_2(m_2)|h_1|^2/|h_2|^2+1\right)}{|h_1|^2} + \frac{m_2 \Gamma_2(m_2)}{|h_2|^2}.
\end{equation}

 Now we have the following Lemma
\vspace{-1mm}
\begin{lem}\label{monotonicity_prop}
Function $m_k\Gamma_k(m_k)$ in \eqref{eq_P_MOMA_target} is strictly decreasing for blocklength $m_k$ satisfying \eqref{p1.2} as long as error probability $\epsilon_k$ and packet size $N_k$ satisfy
\setcounter{equation}{9}
\begin{align} \label{monotonicity_condition}
\frac{Q^{-1}(\epsilon_k)}{\sqrt{N_k}} \le 0.64394\cdots.
\end{align}
\end{lem}
\begin{IEEEproof}
The proof is relegated to Appendix \ref{app1}.
\end{IEEEproof}
It is worthwhile to note that comparing to [6, Proposition 1], condition \eqref{monotonicity_condition} is less restrictive as it allows the monotonicity holds under much milder conditions (e.g., $\epsilon_k \ge 10^{-10}$ and $N_k \ge 100$). Indeed, \eqref{monotonicity_condition} is satisfied in the URLLC system, of which the typically required codeword error probability is $10^{-6}$ and the packet size is around $160$ bits ($20$ bytes) \cite{urllc-ICCW}.
Based on the monotonicity presented in Lemma \ref{monotonicity_prop} , we can obtain the global optimal solution to problem \eqref{P_NOMA1} as follows.

\begin{thm} \label{noma_case1}
Suppose that \eqref{monotonicity_condition} is met and that \eqref{P_NOMA1} is feasible. The optimal solution to problem \eqref{P_NOMA1} is given by
\begin{align} \label{optimal_solution_case1}
\left\{
  \begin{array}{ll}
    m_k^* = D_k,  ~~ {\rm for}~ k=1,2,\\
    \gamma_k^* ~= \Gamma_k(m_k^*),  ~~ {\rm for}~ k=1,2,\\
    p_1^* ~= \frac{\gamma_1^*\gamma_2^*}{|h_2|^2} + \frac{\gamma_1^*}{|h_1|^2},\\
    p_2^* ~= \frac{\gamma_2^*}{|h_2|^2},
  \end{array}
\right.
\end{align}
where optimal SINR $\gamma_k^*~(k=1,2)$ can be obtained through Algorithm \ref{bisection}.
\end{thm}

\begin{IEEEproof}
We first claim that \eqref{p1.3} must hold with equality at the optimum, i.e., the optimal blocklengths must be
$m_k^* = D_k$ for all $k=1,2.$	Suppose that this is not true, i.e., $m_k^* < D_k$ for $k=1$ or $k=2$. Then one can further increase $m_k^*$.
According to  \cite{Xu-2016}, $\Gamma_k(m_k)$ is monotonically decreasing with $m_k > 0$.
Since
\[
p_1+p_2=\frac{\Gamma_1(m_1) \Gamma_2(m_2)}{|h_2|^2}+\frac{\Gamma_1(m_1)}{|h_1|^2}+\frac{ \Gamma_2(m_2)}{|h_2|^2},
\]
$p_1^*+p_2^*$ can be reduced without violating \eqref{p1.4} when $m_k^*$ increases.
Besides, by Lemma \ref{monotonicity_prop}, we also know that $m_k\Gamma_k(m_k)$ is decreasing with $m_k$.
Thus the energy function in \eqref{eq_P_MOMA_target} can be reduced when $m_k^*$ increases. These two facts  contract with the optimality of $m_k^*$. So we must have $m_k^* = D_k$ for all $k=1,2.$
Correspondingly, $\gamma_k^*= \Gamma_k(m^*_k)$ from \eqref{eq_Gamma_P_MOMA} and $p_k^*$ can be obtained from $\gamma_k^*$ from \eqref{p1.6} and \eqref{p1.7} accordingly, which lead to  the optimal solution in \eqref{optimal_solution_case1}.

Note that if $p_1^*+p_2^*=\frac{\gamma_1^*\gamma_2^*}{|h_2|^2} + \frac{\gamma_1^*}{|h_1|^2}+\frac{\gamma_2^*}{|h_2|^2}>P_{\max},$ then it implies that \eqref{P_NOMA1} is infeasible.
\end{IEEEproof}	

To evaluate $\gamma_k^* ~= \Gamma_k(m_k^*)$, $k=1,2$, one can utilize the monotonicity property of $\Gamma_k(m_k)$, and search $\gamma_k^*$ in a bisection strategy, as presented in Algorithm \ref{bisection}. Interestingly, the inverse of implicit function $\Gamma^{-1}_k(\gamma_k)$ can be expressed in closed-form as \eqref{gamma_inv}.

\begin{algorithm}[!tb] \small
\caption{Algorithm to find optimal SINR for problem \eqref{P_NOMA1}}\label{bisection}
\begin{algorithmic}[1]
 \STATE {{\bf Given} the initial values $\Gamma_{\ell k} = 0$, $\Gamma_{uk} = P_{\max}|h_k|^2$, and the tolerance $\epsilon_0$.}\\
 \WHILE {$\Gamma_{uk} - \Gamma_{\ell k} > \epsilon_0$}
 \STATE {$\bar{\gamma}_k = \frac{1}{2}(\Gamma_{uk} + \Gamma_{\ell k})$.}
 \STATE {Compute $\bar{m}_k=\Gamma_k^{-1}(\bar{\gamma}_k)$ as }
 	\!\!\!\!\!\! \begin{align}
        &\Bigg[\frac{1}{2\log_2(1\!+\!\bar{\gamma}_k)}\Bigg(\frac{ Q^{-1}(\epsilon_k)}{\ln 2}\sqrt{1\!-\!\frac{1}{(\bar{\gamma}_k\!+\!1)^2}}\notag \\ &\!\!\!+\!\sqrt{\!\!\left(\!1\!-\!\frac{1}{(\bar{\gamma}_k\!+\!1)^2}\!\right)\!\!\left(\!\frac{Q^{-1}(\epsilon_k)}{\ln 2}\!\right)^2 \!\!\!+\! 4N_k \log_2 (1\!+\!\bar{\gamma}_k)}\!\Bigg) \!\Bigg]^2 \label{gamma_inv}
    \end{align}
 \STATE {{\bf if} $\bar{m}_k < m_k^*$ {\bf then}}\\
 \STATE {\quad Update $\Gamma_{uk} = \bar{\gamma}_k$.}\\
 \STATE {{\bf else}}\\
 \STATE {\quad Update $\Gamma_{\ell k} = \bar{\gamma}_k$.}\\
 \STATE {{\bf end if}}\\
 \ENDWHILE
 \ENSURE {$ \gamma^*_k = \Gamma_k(m_k^*)$}
\end{algorithmic}
\end{algorithm}

\subsection{ Optimal Solutions for Problem \eqref{noma_case_2} and \eqref{noma_case_21}} \label{subSec_solution2}
{Similar to problem \eqref{P_NOMA1}, the optimal solutions of problem \eqref{noma_case_2} and \eqref{noma_case_21} can be obtained by using the monotonicity in Lemma \ref{monotonicity_prop}, which are summarized in the following corollaries.}
\begin{cor} \label{cor_noma_case21}
If condition \eqref{monotonicity_condition} is met, the optimal solution of problem \eqref{noma_case_2} is given by \eqref{optimal_solution_case2} whenever it is feasible.
\begin{align}\label{optimal_solution_case2}
\left\{
  \begin{array}{ll}
    m_k^* = D_k,  ~~ {\rm for}~ k=1,2,\\
    \gamma_k^* ~= \Gamma_k(m_k^*),  ~~ {\rm for}~ k=1,2,\\
    p_1^* ~= \frac{\gamma_1^* |h_2|^2 + \gamma_1^* \gamma_2^* |h_1|^2 }{|h_1|^2|h_2|^2(1 - \gamma_1^* \gamma_2^*)},\\
    p_2^* ~= \frac{\gamma_2^* |h_1|^2 + \gamma_1^* \gamma_2^* |h_2|^2 }{|h_1|^2|h_2|^2(1 - \gamma_1^* \gamma_2^*)},
  \end{array}
\right.
\end{align}
where optimal SINR $\gamma_k^*~ (k=1,2)$ can be obtained through Algorithm \ref{bisection}.
\end{cor}

\begin{cor} \label{cor_noma_case22}
If condition \eqref{monotonicity_condition} is met, the optimal solution of problem \eqref{noma_case_21} is given by \eqref{optimal_solution_case21} whenever it is feasible.
\begin{align} \label{optimal_solution_case21}
\left\{
  \begin{array}{ll}
    m_1^* = m_2^* = D_1,  ~~ {\rm for}~ k=1,2,\\
    \gamma_k^* ~= \Gamma_k(D_1),  ~~ {\rm for}~ k=1,2,\\
    p_1^* ~= \frac{\gamma_1^*}{|h_1|^2},\\
    p_2^* ~= \frac{\gamma_1^* \gamma_2^*}{|h_1|^2} + \frac{\gamma_2^*}{|h_2|^2},
  \end{array}
\right.
\end{align}
where optimal SINR $\gamma_k^*~ (k=1,2)$ can be obtained through Algorithm \ref{bisection}.
\end{cor}

\begin{IEEEproof}
The proofs of Corollary \ref{cor_noma_case21} and \ref{cor_noma_case22} are similar to that of Theorem \ref{noma_case1}. Thus we omit it here.
\end{IEEEproof}

\section{Simulation Results}

\begin{figure}[!tp]
	\centering
	\includegraphics[width=0.96\linewidth]{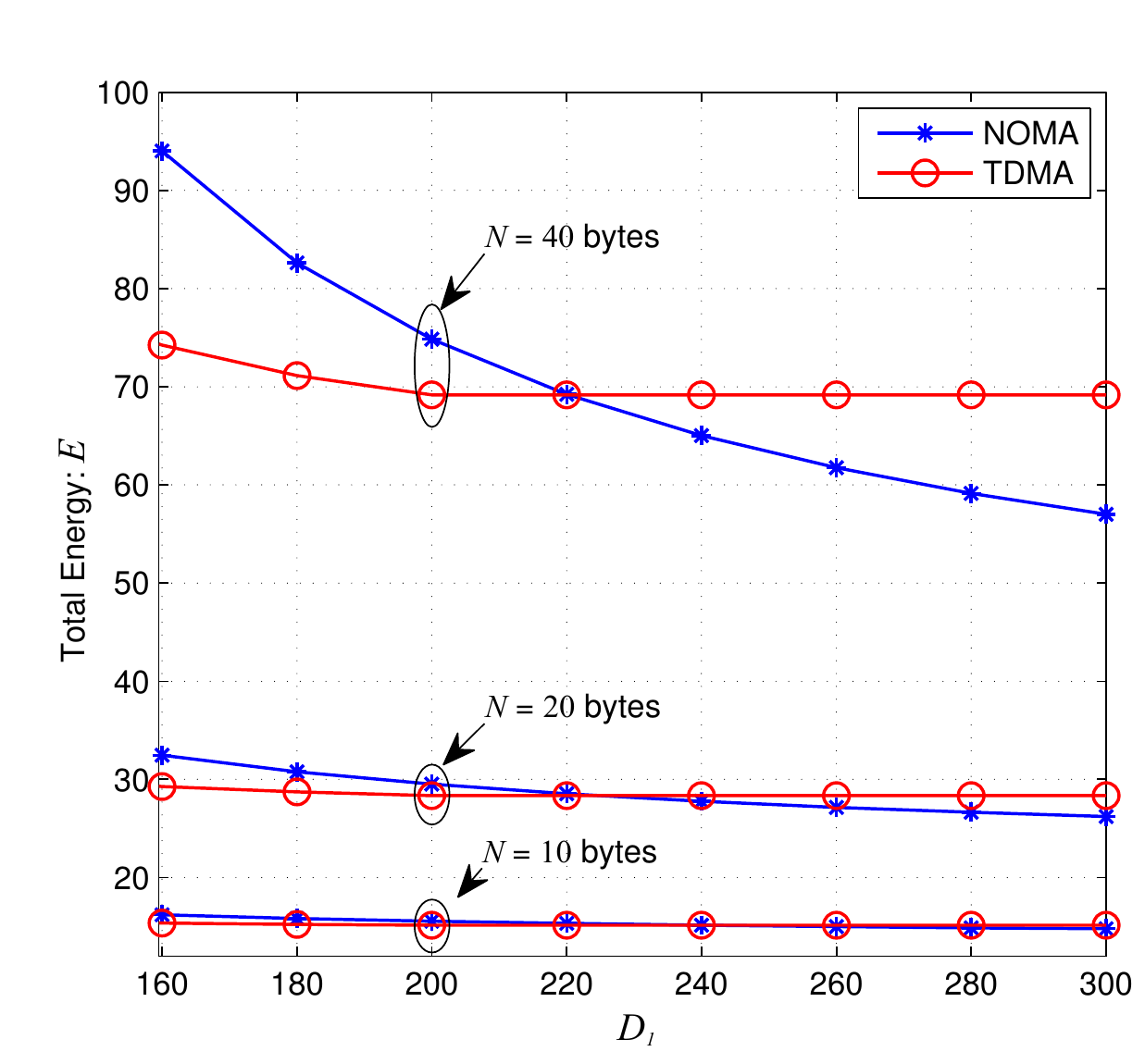}\\
	\caption{A comparison of total consumed energy under different transmission schemes with $D_2 = 300$ and $P_{\max} = 30$ dBm.} \label{energy_comp1}
\end{figure}

\begin{figure}[!tp]
	\centering
	\includegraphics[width=0.96\linewidth]{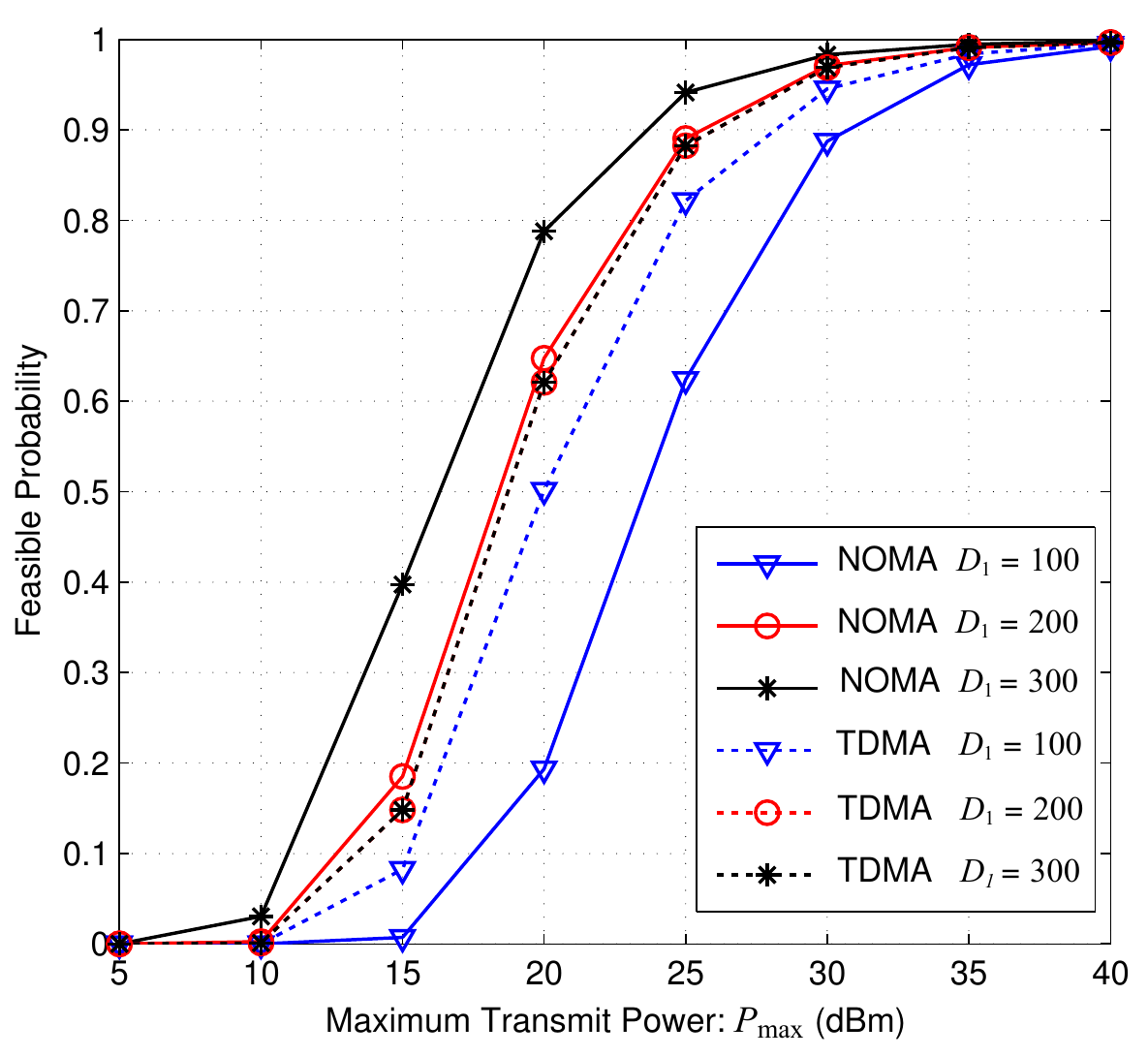}\\
	\caption{Feasible probability comparison of different transmission schemes with $D_2 = 300$.} \label{FP}
	\vspace{-5mm}
\end{figure}


\begin{figure*}[!ht]
\begin{subequations}\label{first_derivative}
\begin{align}
\frac{d E}{d m} &\circeq x F_{x}^{\prime} - mF_m^{\prime} = \frac{mx}{x+1} - \frac{Q \sqrt{m}x}{(x+1)^2 \sqrt{x(x+2)}} - m\ln(x+1) + \frac{Q\sqrt{m}}{2} \frac{\sqrt{x(x+2)}}{x+1}\\
&= \left(\frac{x}{x+1} -\ln(x+1)\right)m + \frac{(x+1)(x(x+2)) - 2x}{2(x+1)^2\sqrt{x(x+2)}}Q\sqrt{m}\\
&\circeq \left(\frac{x}{x+1} -\ln(x+1)\right){\sqrt{m}} + \frac{x^3+3x^2}{2(x+1)^2\sqrt{x(x+2)}}Q   \label{deriv_1}\\
&= \left(\frac{x}{x+1} -\ln(x+1)\right)\frac{\sqrt{x(x+2)}Q + c}{2(x+1)\ln(x+1)} + \frac{x^3+3x^2}{2(x+1)^2\sqrt{x(x+2)}}Q\\
&= \frac{x\sqrt{x(x+2)} Q + cx}{2(x+1)^2 \ln(x+1)} - \frac{\ln(x+1)\sqrt{x(x+2)}Q + c\ln(x+1)}{2(x+1) \ln(x+1)} + \frac{x^3+3x^2}{2(x+1)^2\sqrt{x(x+2)}} Q\\
&\circeq x^2(x\!+\!2)Q \!+\! cx\sqrt{x(x\!+\!2)}\!-\! x(x\!+\!1)(x\!+\!2)\ln(x\!+\!1)Q - c(x\!+\!1)\sqrt{x(x\!+\!2)}\ln(x\!+\!1)+ (x^3+3x^2)\ln(x+1)Q \label{deriv_2}\\
&= \big(x^2(x\!+\!2)\!-\!(x^3\!+\!3x^2\!+\!2x)\ln(x\!+\!1)\!+\!(x^3\!+\!3x^2)\ln(x\!+\!1) \big)Q \!+\! \big(x\!-\!(x\!+\!1)\ln(x\!+\!1)\big)\sqrt{x(x\!+\!2)}c\\
&= \big(x^2(x\!+\!2) \!-\! 2x\ln(x\!+\!1)\big) Q \!+\!\! \big(x\!-\!(x\!+\!1)\ln(x\!+\!1)\big)\!\sqrt{x(x\!+\!2)}\!\sqrt{x(x\!+\!2)Q^2 \!+\! 4N \!(x\!+\!1)^2 \!\ln(x\!+\!1)\!\ln 2}\\
&\le \big(x^2(x\!+\!2)\! -\! 2x\ln(x\!+\!1)\big) Q \!+\! \frac{\sqrt{2}}{2}\big(x\!-\!(x\!+\!1)\ln(x\!+\!1)\big)\sqrt{x(x\!+\!2)}\left(\!\!\sqrt{x(x\!+\!2)Q^2} + \sqrt{4N (x\!+\!1)^2 \ln(x\!+\!1)\ln 2}\right) \label{deriv_3}\\
&\le \left(\!\!x^2(x\!+\!2) \!-\! 2x\frac{2x}{x\!+\!2}\!\!\right)\!Q  \!+\!\frac{\sqrt{2}}{2}\!\!\left(\!\!x\!-\!(x\!+\!1)\frac{2x}{x\!+\!2}\!\!\right)\!x(x\!+\!2)Q \!+ \!\sqrt{2}\!\left(\!x\!-\!(x\!+\!1)\frac{2x}{x\!+\!2}\!\right)\!(x\!+\!1)\! \sqrt{x(x\!+\!2)} \sqrt{\frac{2x}{x\!+\!2}N\ln 2} \label{deriv_4}\\
&= \left(x^2(x+2)^2 - 4x^2\right)(x+2) Q - \frac{\sqrt{2}}{2}x^3(x+2)^2Q - \sqrt{2}x^2(x+1) \sqrt{x(x+2)} \sqrt{2x(x+2)N\ln 2}\\
&= x^3(x+4)(x+2) Q - \frac{\sqrt{2}}{2}x^3(x+2)^2Q - 2 x^3(x+1) (x+2) \sqrt{N\ln 2}\\
&\circeq 2(x+4) Q - \sqrt{2}(x+2)Q - 4(x+1) \sqrt{N\ln 2}   \label{deriv_5}\\
&= \underbrace{(2Q-\sqrt{2}Q - 4\sqrt{N\ln 2})}_{f_1(Q,N)}x + \underbrace{8Q-2\sqrt{2} Q -4 \sqrt{N\ln 2}}_{f_2(Q,N)}.
\end{align}
\end{subequations}
\hrulefill
\end{figure*}

In this section, simulation results are given to compare the performance of NOMA with that of the time-division multiple access (TDMA) under FBC. We assume that the packets contain equal size of $20$ bytes; the block error probability are set to be $\epsilon_1 = \epsilon_2 = 10^{-7}$, and that the channel coefficients are modelled by Rayleigh distribution with a scale parameter equals to $100$. When $|h_1|>|h_2|$, both problems  \eqref{noma_case_2} and \eqref{noma_case_21} are solved and the one that yields the smaller energy  is chosen as the consumed energy of the NOMA scheme. The energy of TDMA is solved from the SUM method in \cite{Xu-2016}.

As shown in Fig. \ref{energy_comp1}, the total consumed energies for both schemes decline with the increase of $D_1$. Specifically, for the NOMA scheme, the total consumed energy is strictly decreasing with $D_1$. The reason is that
a lager $D_1$ indicates a larger $m_1$ and  a smaller energy consumption for delivering the packet of receiver $1$.
The consumed energy of receiver $2$ is also decreased due to the reduced interference from receiver $1$. Therefore the total consumed energy in NOMA scheme decreases strictly with the increase of $D_1$.
While for the TDMA scheme, the whole transmission time is dominated by $D_2$. When $D_1$ is small, the optimal $m_1$ satisfies $m_1^* = D_1$. When $D_1$ increases, the total consumed energy decreases. When $D_1$ is large enough, the optimal $m_1$ satisfies $m_1^* < D_1$, and $m_1^*$ keeps constant with the increase of $D_1$. As a result, the total consumed energy decreases first and then keeps invariant in TDMA scheme. 
It is important to note that the NOMA scheme gradually performs better than the TDMA scheme as $D_1$ approaches $D_2$ since SIC can effectively reduce the interference as long as it is feasible.

Fig. \ref{FP} depicts that probability for at least one of the energy-minimization problems being feasible, for different values of
$D_1$ and $P_{\max}$. Specifically,  one can see that the feasible probability of the NOMA scheme increases with the increase of $D_1$, while the feasible probabilities of the TDMA scheme with $D_1 = 200$ and $D_1 = 300$ are very close, implying that in most channel realizations the optimal $m^*_1 \le 200$, thus increasing $D_1$ has a slight effect on the feasible probability of the problem.

\vspace{-2mm}
\section{Conclusions}
In this paper, based on the FBC capacity formula, we have considered the energy-efficient NOMA transmission design problem subject to heterogeneous latency and reliability constraints.
In particular, by considering different channel conditions and latency constraints, we have formulated three NOMA design problems in \eqref{P_NOMA1}, \eqref{noma_case_2} and \eqref{noma_case_21}, respectively.
By the monotonicity of the energy function with respect to the blocklength, we have shown that the three problems admit simple closed-form solutions. The presented numerical results have also shown that the NOMA schemes can be more  favourable than the TDMA scheme if the receivers have similar latency requirements.
%
%

\begin{appendices}
\section{Proof of Lemma \ref{monotonicity_prop} }\label{app1}
For the simplicity of notation, we remove the subindex of all variables and let $x$ denote the SINR.
Based on \eqref{p1.1}, we define
\begin{align}\label{func_F}
F(m,x) &\triangleq m\ln(x + 1) - \sqrt{m} \frac{\sqrt{x(x+2)}}{(x + 1)} Q^{-1}(\epsilon) - N \ln2 \notag\\
       &= 0.
\end{align}

Note that the right-hand side of \eqref{func_F} is a quadratic equation of $\sqrt{m}$. By letting $Q = Q^{-1}(\epsilon)$ and $c = \sqrt{x(x+2)Q^2 + 4(x+1)^2 \ln(x+1) N \ln 2 }$, the positive root of \eqref{func_F} is given by
\begin{align}
\sqrt{m} = \frac{\sqrt{x(x+2)}Q + c}{2(x+1)\ln(x+1)}.
\end{align}
According to the implicit function theorem, the partial derivatives of $F(m,x)$ with $m$ and $x$ can be described as
\begin{subequations}
\begin{align}
F_m^{\prime} &= \ln(x+1) - \frac{Q}{2\sqrt{m}} \frac{\sqrt{x(x+2)}}{x+1},\\
F_{x}^{\prime} &= \frac{m}{x+1} - \frac{Q \sqrt{m}}{(x+1)^2 \sqrt{x(x+2)}}.
\end{align}
\end{subequations}
As is shown in \cite{Xu-2016}, $F_m^{\prime}>0$ and $F_{x}^{\prime}>0$ always hold  with $m>0$ and $x>0$ respectively. Thus the monotonicity of $\bar{E}(m) = m_k \Gamma_k(m_k)$ can be verified by
\begin{align}
\frac{d \bar{E}}{d m} \!=\! \frac{d mx}{d m} \!=\! x \!+\! m\frac{d x}{d m} \!=\! x \!-\! m\frac{F_m^{\prime}}{F_{x}^{\prime}} \!\circeq\! x F_{x}^{\prime} - mF_m^{\prime},
\end{align}
where $A \circeq B$ denotes that $A$ and $B$ have the same sign. The first-order derivative of $\bar{E}(m)$ is given by \eqref{first_derivative}, where \eqref{deriv_1} and \eqref{deriv_2} hold due to $\sqrt{m} >0$ and $2\sqrt{x(x+2)}(x+1)^2\ln(x+1)>0$ for $x>0$ respectively. \eqref{deriv_3} holds because of $x-(x+1)\ln(x+1)<0$ with $x>0$ and the fact that $2\sqrt{a+b} \ge \sqrt{2}\big(\sqrt{a}+\sqrt{b}\big)$ for $a,b>0$. \eqref{deriv_4} holds owing to $\ln(x+1)\ge \frac{2x}{x+2}$ for $x>0$. In addition, \eqref{deriv_5} holds since $x^3(x+2) > 0$ for $x>0$.

To prove that $\bar{E}(m)$ is a monotonically decreasing function, we need
$f_1(Q,N)<0$ and $f_2(Q,N)<0$, indicating $\frac{Q}{\sqrt{N}} \le \frac{2\sqrt{\ln2}}{4-\sqrt{2}} = 0.64394\cdots$.
Noting that $f_1(Q,N)$ and $f_2(Q,N)$ increase with $Q$ and decrease with $N$, and $Q = Q^{-1}(\epsilon)$ is a monotonically decreasing function with $\epsilon$, therefore, for any larger $\epsilon$ and $N$, the monotonicity of $\bar{E}(m)$ hold. This completes the proof.
\hfill $\blacksquare$

\end{appendices}

{
\smaller[1]

}

\end{document}